\documentclass[10pt, a4paper]{article}

\usepackage[final]{lrec2026} 

\usepackage{adjustbox}
\usepackage{tipa}
\usepackage{amsmath} 

\usepackage{multirow}
\usepackage{hhline}
\usepackage{makecell}
\usepackage{booktabs}
\usepackage{array}
\usepackage{caption}

\usepackage{graphicx}

\title{spINAch: A Diachronic Corpus of French Broadcast Speech Controlled for Speakers' Age and Gender} 

\name{Simon Devauchelle$^{\ast\dagger}$, David Doukhan$^{\dagger}$, Rémi Uro$^{\ddag}$, Lucas Ondel Yang$^{\ast}$, \\ 
{\bf \large Valentin Pelloin$^{\dagger}$,
Olympia Imbert-Brégégère$^{\dagger}$, Véronique Lefort$^{\dagger}$,} \\ {\bf \large Kévin Picard$^{\dagger}$, Emeline Seignobos$^{\dagger}$, Albert Rilliard$^{\ast}$} }

\address{\\$^{\ast}$Université Paris Saclay, CNRS, LISN - Orsay, France;\\
$^{\dagger}$Institut National de l'Audiovisuel - Paris, France;\\
$^{\ddag}$LIASD, Université Paris 8 - Saint-Denis, France.\\
\\
         simon.devauchelle@universite-paris-saclay.fr, 
         r.uro@iut.univ-paris8.fr,
         lucas.ondel@cnrs.fr, \\
         \{ddoukhan,vpelloin,olimbertbregegere,vlefort,kpicard,eseignobos\}@ina.fr,
         albert.rilliard@lisn.fr\\
         }

\abstract{
We present \textit{spINAch}, a large diachronic corpus of French speech from radio and television archives, balanced by speakers' gender, age (20-95 years old), and spanning 60 years from 1955 to 2015. 
The dataset includes over 320 hours of recordings from more than two thousand speakers.
The methodology for building the corpus is described, focusing on the quality of collected samples in acoustic terms. 
The data were automatically transcribed and phonetically aligned to allow studies at a phonemic level. 
More than 3 million oral vowels have been analyzed to propose their fundamental frequency and formants.
The corpus, available to the community for research purposes, is valuable for describing the evolution of Parisian French through the representation of gender and age. 
The presented analyses also demonstrate that the diachronic nature of the corpus allows the observation of various phonetic phenomena, such as the evolution of voice pitch over time (which does not differ by gender in our data) and the neutralization of the \textipa{/a/}-\textipa{/A/} opposition in Parisian French during this period.
\newline \Keywords{
    Speech Corpus,  
    Diachrony,
    Broadcast News, 
    Parisian French, 
    Gender and Age Bias evaluation} 
 }

\begin{document}

\maketitleabstract

\section{Introduction}

Diachronic changes in speech may be studied longitudinally, focusing on a specific speaker who may represent a given population category, as in \citet{Harrington_Palethorpe_Watson_2000}. 
Other longitudinal studies are more focused on individual changes in relation to specific and documented life events \cite{Riverin-Coutlee_Harrington_2022}.
Cross-sectional corpora, like the one described by \citet{Stuart-Smith_2020}, allow for investigating changes at the population level. \citet{Stuart-Smith_2020} stratified their speakers according to social gender and age (middle-aged vs. younger) with two dates of recordings, allowing them to study language changes over four levels of date of birth, and introducing a distinction between real- and apparent-time differences. 
Real-time differences are found with the two dates of recording, and apparent-time differences are obtained by factoring in these recording dates with the actual ages of the speaker to consider the dates of birth. 
Studying speech variation across time raises the notable challenge of finding old recordings that can be compared to more recent ones. 
The scarcity of such resources explains why some studies compare only two groups of speakers -- for example, one from the 1940s and one from the 1990s in \citet{Pemberton_McCormack_Russell_1998}.
Another important question when selecting a speech dataset is related to the speech style(s) it represents, with read vs. spontaneous speech being known to induce differences at various levels \cite{Hollien_Hollien_DeJong_1997}.

The way we speak and its evolution over time is influenced by many factors, including contacts between populations \cite{Mufwene_2007}, or the identification of social groups to shared reference from media outlets that can shape identity display \cite{Stuart-Smith_2006, Vigouroux_2015}.
Gathering resources capable of some generalization over a population requires a large and complex dataset.
While corpora such as $VoxCeleb$ \citeplanguageresource{Nagrani_Chung_Zisserman_2017} or $Common Voice$ \citeplanguageresource{ardila-etal-2020-common} feature thousands of speakers and very large acoustic datasets (over 100k hours), they are mostly synchronic resources and thus not well suited for studying language evolution.
Large resources in terms of speaker diversity are a rare feature within diachronic corpora described in the literature (\citealp[e.g.,][ is a relatively large resource, but features very few speakers]{Zou_Wang_He_2012}).
Typical diachronic datasets feature dozens of speakers (\citealp[e.g.,][]{Pemberton_McCormack_Russell_1998, Stuart-Smith_2020}), and are generally based on read speech, with notable exceptions \cite{Hollien_Green_Massey_1994, Barras_Allauzen_Lamel_Gauvain_2002}.
Two diachronic datasets based on broadcast archives were recently described for French \cite{Suire_Barkat-Defradas_2020, uro22}, and feature hundreds of speakers, with some gender balance, but have not been made available to the community due to authorship considerations, copyright restrictions, privacy concerns, etc.

In this paper, we present and analyze a new large-scale cross-sectional corpus of French, \textit{spINAch}, which is made freely available to the research community. 
The complete corpus is freely available for research purposes at  \url{https://www.ina.fr/institut-national-audiovisuel/research/dataset-project#spINAch}. 
The acoustics estimates presented in Section~\ref{sec:acoustic} are directly available at \url{https://doi.org/10.5281/zenodo.18714702}.
This corpus comprises audio recordings of more than 2,000 speakers, recorded over a sixty-year-long time span (from the 1950s to the 2010s).
The data is composed of excerpts from French radio and television archives from the \textit{Institut National de l'Audiovisuel} (INA). 
Archivists prepared a list of potential speakers participating in broadcast shows (focusing on interviews and talk shows) in order to target known individuals. 
This allowed a stratification in terms of speakers' Age (between 20 and 95 years old) and Gender, for seven time Periods (a 10-year time span was selected between 1955 and 2015).
The recordings, totaling more than 320 hours of speech, were automatically transcribed and forced aligned in order to allow the extraction of acoustic analyses (formants and fundamental frequency, $f_o$). 
This dataset (including the audio recordings, their automatic and manual transcriptions, acoustic analyses, with anonymous demographic information about the speakers) is made available to the research community. 
We present in section~\ref{sec:corpus} the methods used to gather and analyze this large dataset, with details on its composition, the acoustic measurements made, and quality evaluations. Section~\ref{sec:analysis} proposes some preliminary diachronic analysis of the changes that can be observed across this time span for French spoken in national media outlets.

\section{Corpus description}
\label{sec:corpus}

The corpus was collected in two iterations (\citealp[the first being described in][]{uro22}) with an identical methodology, except for
improvements of state-of-the-art diarization and music detection methods.
An evaluation of the extraction methods of the two phases applied on a subsample is given in section~\ref{sec:evalex} to ascertain that the resulting acoustic segments propose comparable linguistic information, if obtained through different processes.

\subsection{Archive selection}

An essential step in building a gender- and age-balanced cross-sectional corpus over sixty years was to spot in the archives' metadata potential target speakers that match the Age and Gender criteria: female and male speakers spread across four age groups (20-34, 35-49, 50-64, and over 64 years old), in equivalent number at seven time Periods separated by 10 year time-steps (1955-1956, 1965-1956, 1975-1976, 1985-1986, 1995-1996, 2005-2006, 2015-2016).  
A target of 30 speakers per Age, Gender, and Period category was set, without duplicates across categories.
The construction of such a balanced corpus was only possible thanks to the expertise of INA's archivists, who parsed the television and radio databases to identify potential target speakers. For each period, they selected media with reasonable acoustic quality, such as studio-recorded talk shows free of background noise featuring interactive conversations, and verified that participants had enough speaking time. Achieving this requires archivists to review or listen to the collections. By cross-referencing the speaker's birth date with the date of the first broadcast, they estimated the participant's age. The compilation of this corpus involved a back-and-forth process with the archivists’ identification work, which helped us fill in the missing speaker categories. 
Based on INA's documentation databases, archivists identified about 10,000 individuals who matched these characteristics. 
Difficulties in completing some profiles could not always be overcome, especially for women, and for younger or older persons, from the earliest periods (see Table~\ref{tab:corpus}). This bias of female representation in the media is known and well documented \cite{CoulombGully_2011, Doukhan_2018B}.

\subsection{Sound-Track extraction}

Automated signal-processing routines were applied to obtain single-track WAV files sampled at 16~kHz from archives, and to discard recordings having undesirable properties.
A first decompression step was performed using \texttt{ffmpeg}, leading to up to 5 uncompressed tracks (stereo, mono, or Dolby) from heterogeneous archives encoded with various codecs.
Audio tracks were inspected for the presence of speaking clock: a method used from 1970 to 1990 to embed time-code information in archives using a dedicated audio track~\cite{vallet14}.
The speaking clock was spotted using a 1000~Hz beep detector, along with hard-coded rules describing characteristics of its temporal patterns, leading to the exclusion of the corresponding audio tracks.
A signal bandwidth estimator, based on the cumulative sum of the long-term spectrogram, was used to discard recordings with bandwidth below 8~kHz, often corresponding to undesirable archive encoding or transcoding strategies that may result in biased acoustic parameter extraction.
Lastly, we used autocorrelation to detect time delays between the remaining audio tracks.
When a time delay was detected, we kept only the first track; otherwise, we mixed the remaining tracks.

\subsection{Manual speaker identification}

The next step, and the more time-consuming one, was to identify whether and when each targeted speaker actually speaks within the raw audio archives.
This was a manual process, supported by the voice activity detection (VAD) and the diarization of each archive.
Speaker identification was realized by four authors of this study, resulting in a total involvement of about 40 days.
Preprocessed audiovisual archives were presented to annotators using ELAN~\cite{sloetjes2008annotation}, displaying the cluster identifiers obtained during the diarization and cleaning process (see section \ref{sec:data_extraction_cleaning}), synchronized with the archives' audio and video tracks.
Annotators were provided with a shared spreadsheet containing a list of archive identifiers and target speakers.
The list was enriched with details about the target speaker (gender, age,  occupation) to support the identification process.
Annotators reported the target speaker's cluster identifier in the spreadsheet.

Several criteria were defined along the identification process to reject speakers having undesirable properties, resulting in a manual rejection rate of about 11\%: bad acoustic quality of speaker utterances (telephone speech, outdoor recordings, large amounts of background noise or music, strong audio effects), diarization under-segmentation errors resulting in several speakers sharing the same cluster identifier, use of foreign language or strong non-French speaking accent, dubbed speaker, homonym speaker having different age and occupation than the target, retrospective show broadcasting the voice of the target speaker over several decades resulting in incorrect speaker age estimation, etc.

\subsection{Data extraction, cleaning, and transcription}
\label{sec:data_extraction_cleaning}
Once a target speaker is identified, all segments obtained from \textit{pyannote} [v3.1] \cite{bredin23_interspeech} diarization corresponding to their voice are extracted, excluding overlapping voice segments.
The audio excerpts are submitted to a cleaning procedure because the primary objective of the corpus is studying speech's acoustic characteristics -- a process sensitive to the presence of noise or background music. On top of \textit{pyannote} predictions, we also applied \textit{InaSpeechSegmenter} (\citealp[$ISS$ v0.8;][]{Doukhan_2018}) as a voice activity detector and merged its outputs with those from the diarization in order to better identify spoken segments. Outputs from \textit{pyannote} overlapping (even marginally) non-speech events detected by $ISS$ were discarded. To avoid any acoustic bias from telephone-quality speech segments, we use the \textit{LIUM~SpkDiarization} [v8.4.1] \cite{Meignier_2010} to detect and remove them. 
Some cleaning was already done when applying \textit{pyannote} and $ISS$, as it removes what is considered noise or music to keep only speech, but background music is still possible. We use a music detection model and apply a threshold of 0.8 to the ratio of the detected duration to the segment's total duration.
The music segmentation model\footnote{\scriptsize \url{https://hf.co/ina-foss/ssl-music-detection-music2vec}}
 described in \citet{pelloin_lrec2026} uses \textit{music2vec} \cite{li2022mapmusic2vecsimpleeffectivebaseline} embeddings and classifies each frame. It obtains a frame-level F1-Score of 89.7\% on \textit{OpenBMAT} \citeplanguageresource{MelndezCataln2019} and 92.0\% on \textit{Seyerlehner} \citeplanguageresource{seyerlehner2007automatic}, two music detection datasets of TV broadcast content.
At the end of the cleaning process, after removing all parts containing noise, music, or overlapping speech, we obtained the diarized speech segments for each target speaker.

These speech excerpts were then fed to \textit{Whisper} [large-v3] \cite{Radford_2022} in order to obtain a lexical transcription.
This transcription was used to force-align its phonetic transcription to the speech signal using the \textit{Montreal Forced Aligner} (\citealp[$MFA$, version 3.0;][]{McAuliffe_Socolof_Mihuc_Wagner_Sonderegger_2017}).
An evaluation of the transcription accuracy, both in terms of word error rate and of phone error rate, is given in section~\ref{sec:transcrip}. Without applying any cleaning (i.e., using raw segments from \textit{pyannote} and $ISS$), the total number of phones from the $MFA$ output exceeded 4.6M oral vowels. Vowels with predicted duration over 200 milliseconds were removed (about 5.1\%). After removing unvoiced vowels using the method described in the next section \ref{sec:acoustic}, the music detection model reduces again this vowel set by 8.2\%.
This leads to a set of 3,016,134 million vowels available in the clean version of the corpus.

\subsection{Acoustic measurements}
\label{sec:acoustic}

Several acoustic parameters have been extracted and are provided with the corpus to support phonetic analyses of the speech productions it contains.
First, the speech's $f_o$ was estimated for each segment with a 10~ms time step, using two different pitch detection algorithms for robustness \cite{Vaysse_Astesano_Farinas_2022}: the autocorrelation algorithm implemented in \textit{Praat} \cite{Boersma_1993, Boersma_Weenink_2025}, and the \textit{REAPER} estimator \cite{Talkin_2015}.
Frames that any of the two algorithms annotated as unvoiced were deemed unvoiced, and frames where they differ by more than a 20\% gross-error difference in their $f_o$ prediction were also marked as unvoiced -- because possibly unreliable.
For the remaining frames (about 79\% of the total number of frames), the value estimated by $Praat$ was retained.

Then, the first five formants were estimated along the signal using \textit{Praat}'s implementation of the Burg algorithm, with the same 10~ms time step, and adapting the ceiling parameter for each speaker and vowel category following the strategy proposed by \citet{Escudero_Boersma_Rauber_Bion_2009}.
The strategy consists of estimating formants for a set of ceilings, and using the one that minimizes the formants' variance for a given vowel category and a specific speaker.
In our case, we used a set of twenty ceilings above and below the reference ceiling recommended for female and male speakers in $Praat$ documentation (respectively 5.5~kHz and 5~kHz), spaced by steps of approximately 50~Hz (for details, see \textit{Praat}'s documentation\footnote{\scriptsize \url{https://www.fon.hum.uva.nl/praat/manual/FormantPath.html}}) above or below the reference ceiling.
The best ceiling that was kept minimizes the sum of variances observed for the first three formants of all the vowels of a given category for a specific speaker. 
We used the first three formants, in place of the first two in \citet{Escudero_Boersma_Rauber_Bion_2009} because the third formant is relevant for rounding, which is an important feature of the French vocalic system (\citealp[e.g.,][]{Menard_Dupont_Baum_Aubin_2009}).

For each formant, the median of all values observed along the middle third of each vowel was kept.
Formants are expressed in Hertz and converted to a Bark scale using the equation in \citet{Traunmuller_1990}.
For $f_o$, the median of all valid values observed along the vowel was considered.
The $f_o$ values are expressed in Hertz and in semitones (relative to 1Hz).

\subsection{Quality evaluations}

\subsubsection{Evaluation of automatic transcription and forced alignment}
\label{sec:transcrip}

A crucial aspect of this corpus construction lies in its fully automatic transcription and subsequent phonetic alignment, which enable the study of the acoustic characteristics of vowels over time.
To evaluate this process, one hour of speech was randomly sampled from the total corpus, representing 245 speakers spread across the seven periods.
These samples were manually transcribed by four L1 French speakers to reflect their full content of speech. These human-made transcriptions were then submitted to the same $MFA$-based forced-alignment to obtain a phonetic version. Primary concerns related to the transcription quality were related to the possibility of $Whisper$ adding lexical items that were not actually pronounced, a risk that may be increased in older recordings --~acoustically uncommon~-- and therefore less likely to have been included in $Whisper$’s training corpus.
The automatic transcription quality is evaluated using word error rate ($WER$) and phone error rate ($PER$). 

After the text normalization, the $WER$ is 11.7 using \textit{Whisper} large-v3. This score is one point higher than that reported for \textit{CommonVoice}~\text{15} \cite{ardila-etal-2020-common} on the French subset. The phone-level evaluation is a crucial component for the acoustic analysis of the vowels detailed in section \ref{sec:analysis}, given the nature of our dataset (recorded speech from interviews in which speakers may repeat themselves and exhibit disfluencies or hesitations).
The results indicate that the $PER$ (substitutions, deletions, and insertions summed and divided by the total number of phones) reached 7.74, with a phone-level precision of 93.7\%, discarding the risk of major insertions. When focusing only on vowels, the $PER$ degraded by about three points, reaching 10.26. The most common errors made by \textit{Whisper} were deletions. They represent 64.39\% of phone-level errors. On the evaluation subset, 5.45\% of vowels were deleted during the automatic alignment process, compared to the manual transcriptions. The first three most deleted vowels are the \textipa{/\oe/} (58\%), \textipa{/ø/} (22\%) followed by the \textipa{/\textschwa/} (9\%). Hesitations, expressed in French by words like "heu" or "euh" (phonetized by \textipa{/\oe/} or \textipa{/ø/}), represent 28\% of the deleted vowels. Then, the next 5\% are derived from the conjunction "et" (\textipa{/e/}), followed by the words "de" and "que" (4.9\% and 3.5\%) -- these words are likely used in repetitions and hesitations and were only transcribed once by $Whisper$. Among all vowel classes, the vowel \textipa{/\textschwa/} is the most frequently inserted and is mostly derived from the negation word "ne" (23\% of \textipa{/\textschwa/} insertions), which is generally not produced in French \cite{Abeille_Godard_2021}. $Whisper$ tends to add it, resulting in more formal lexical predictions.

\subsubsection{Comparison of extraction methods}
\label{sec:evalex}
Given the size of the corpus, it was produced in two phases (in 2021 and 2025), separated by four years.
The processing for the first version prioritized speech quality when identifying speakers, at the risk of missing potential targets and thereby lowering recall. 
In addition, more up-to-date software has since been released, so the automated processing of archives differs slightly between the two phases, mainly in the algorithms employed (\citealp[for details on the first version, see][]{uro22}).
This notably explains why there are more speakers in the years 1965, 1985, and 2005, as the newer version was more efficient.

In order to evaluate if these two processing methods introduced a qualitative difference in terms of the distribution of the acoustic characteristics of phonemes, an evaluation was necessary, as both processing methods end up with comparable, but not identical, speech segments out of the same original recording.
The main differences introduced by the two processings are (i) different diarization of the archive, and (ii) possibly different transcription and alignment, as the cleaning process (removing background music, noises, etc) was performed with different algorithms.
Our evaluation thus focuses on verifying that the phonetic characteristics of a given speaker (in our case, the distribution of their formants for oral vowels) are comparable across extraction methods. 
That is, the \textipa{/a/}s (and other vowels) extracted with the two methods shall have comparable acoustic characteristics for the same speaker. 
The question here is to verify, for a subset of speakers from the first phase \cite{uro22}, if the formants measured on vowels detected using the new processing are equivalent to those measured on vowels detected by the initial method.
Even if the two methods give slightly different sets of vowels (in terms of the number of vowels observed for a given speaker), on a sufficiently long dataset, the formants of each vowel category shall have comparable distributions within each speaker, across processing methods, if one thinks the extraction method in itself does not bias the phonetic content.

To evaluate the possibility of a bias linked to the processing chain, eight speakers were randomly drawn from the eight categories of the Periods (1955-1956, 1975-1976, 1995-1996, 2015-2016) and Gender (Female, Male) processed at the first phase, for a total of 64 individuals (8 speakers from 8 categories).
The new extraction method was applied to the 64 archives of the initial corpus to extract and transcribe the target speakers' voices a second time.
From these two sets of extracted vowel segments, thanks to the initial (therefore Method [1]) and newer software (therefore Method [2]), the same phoneticization pipeline and the same $f_o$ and formant detection process (see section~\ref{sec:acoustic}) were applied.
An equivalent number of vowels, approximately 90,000, is obtained by both Methods.
Details regarding the number of occurrences obtained for each oral vowel are presented in Table~\ref{tab:compareMethods}.

\begin{table}[!ht]
\begin{center}
\begin{tabular}{crr}
      \toprule
        \textbf{Phone} & \textbf{Method [1]} & \textbf{Method [2]} \\
      \midrule
    \textipa{[i]} & 13,338 & 14,194 \\
    \textipa{[e]} & 13,132 & 13,814 \\
    \textipa{[E]} & 13,483 & 14,317 \\  
    \textipa{[a]} & 18,140 & 19,313 \\
    \textipa{[A]} &    769 &    828 \\
    \textipa{[O]} &  5,930 &  6,332 \\  
    \textipa{[o]} &  2,960 &  3,172 \\
    \textipa{[u]} &  4,146 &  4,511 \\
    \textipa{[y]} &  4,721 &  4,961 \\
    \textipa{[ø]} &  2,652 &  2,799 \\
    \textipa{[@]} &  8,705 &  9,167 \\  
    \textipa{[œ]} &  1,165 &  1,257 \\
      \midrule
      \midrule
Total       & 89,141 &  94,665  \\
      \bottomrule
\end{tabular}
\caption{
Number of occurrences for each vowel category for the two segment sets obtained by the two extraction Methods on the same 64 speakers.}
\label{tab:compareMethods}
\end{center}
\end{table}

A linear mixed model was then fit to the value of each of the first three formants, with the extraction Method ([1]/[2]) as the main predictor, controlled for Gender and Vowel category, and using Periods and Speakers (nested within gender and period) as random variables. This is formalized in formula~\ref{eq:mod1}, that follows R's \textit{lme4} syntax \cite{rcoreteam, bates2015}, and where $z_i$ are the standardized values of either of the first three formants in Bark, $M$ is the \textit{Method} ([1] or [2]) used for processing the archives, $G$ being the \textit{Gender} of a given speaker, $V$ the \textit{Vowel} class (12 levels, see Table~\ref{tab:compareMethods}), $P$ the \textit{Period} (4 levels), and $S$ the index of the \textit{Speaker} producing a given phone:
\begin{equation} \label{eq:mod1}
z_{i} \sim  M + G + V + (1|P/G/S)
\end{equation}
The tables summarizing the factors (fixed and random) of the three regression models (one for each formant) are given in Appendix A (Section~\ref{sec:Appendix_A}).

The models showed there were no significant differences introduced in the distribution of these formants by the extraction \textit{Method} (for $F_1$: $\mathrm{\chi}_{(1)}^{2}=0.084, p=0.772$; for $F_2$: $\mathrm{\chi}_{(1)}^{2}=0.96, p=0.327$; for $F_3$: $\mathrm{\chi}_{(1)}^{2}=0.6147, p=0.433$), while the other parameters had, obviously, major effects on the formants values --- primarily the Vowel categories, but also the gender and a large inter-speaker variability (cf. Tables~\ref{tab:anova_fixed}~\&~\ref{tab:anova_random}).
This result confirms that the extraction method employed here is robust to the evolution of diarization, speech-to-text, forced alignment, and music detection algorithms. 
This is fortunate given the surge of use of comparable frameworks in phonetic studies \citealp[e.g.,][]{Ballier_Meli_2024, Coats_2025, christodoulidou25_interspeech}.
We believe this comparison of the two methods supports that the extracted vowels from all Periods of the corpus shall give coherent information, regardless of how they were segmented.
We consider the data extracted using these two methods to be comparable in terms of phonetic distribution, which justifies prioritizing improvements to the processing pipeline by incorporating state-of-the-art advances in the second phase.

\subsection{Summary of corpus features}

\begin{figure}    \centering\includegraphics[width=\linewidth]{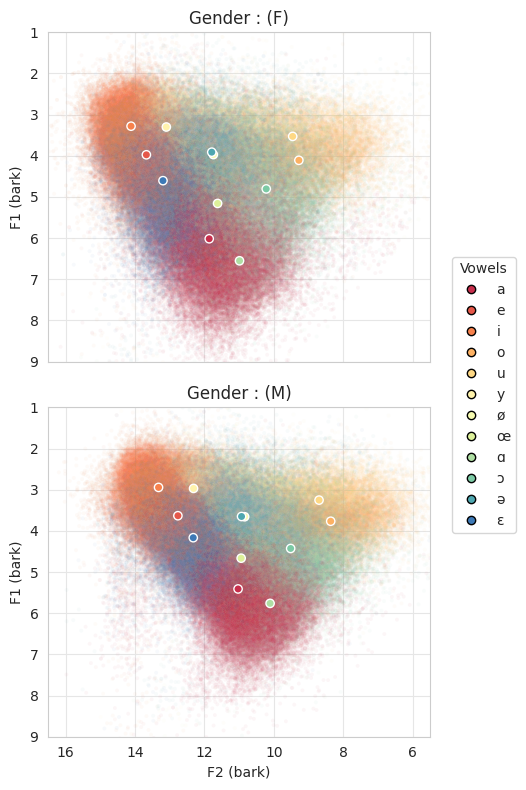}
    \caption{
    Plot of vowels' $F_1$ and $F_2$ (in Bark) values obtained from a sample of 240 speakers in the \textit{spINAch} corpus, balanced in terms of period and gender. Median values by vowel category (encircled in white) and for each vowel (color points) are shown for both genders.}
    \label{fig:triangles}
\end{figure}

Table~\ref{tab:corpus} presents the distribution of the data in terms of speaker count and recording duration across the seven periods, with columns presenting gender and rows presenting age categories. A sample of oral vowel formants estimated using the procedure described above is shown in Figure~\ref{fig:triangles}, resulting in well-known vocalic triangles.
Cleaned oral vowels' distribution is reported in the table in Appendix B \ref{tab:oral_vowels_table}~\footnote{Acoustic estimates from the aligned oral vowels are hosted at \textit{Zenodo} on this url \url{https://doi.org/10.5281/zenodo.18714702}.}.  
Details about the whole phone dataset and its categories (including nasal vowels and consonants) are given in Appendix C (\ref{tab:phones_table}).

\begin{table*}[!ht]
\centering
\scriptsize
\begin{adjustbox}{width=\textwidth}
\begin{tabular}{|c|c|c|c|c|c|c|c|c|c|c|c|c|c|c|c|}
    \cline{2-16}
    \multicolumn{1}{c|}{} &  \multicolumn{2}{c|}{1955/6}  &  \multicolumn{2}{c|}{1965/6}  &  \multicolumn{2}{c|}{1975/6}  &  \multicolumn{2}{c|}{1985/6}  &  \multicolumn{2}{c|}{1995/6}  &   \multicolumn{2}{c|}{2005/6}  &  \multicolumn{2}{c|}{2015/6}  & \multirow{2}{*}{Total} \\
    \cline{2-15}
    \multicolumn{1}{c|}{} & F & M  & F & M & F & M & F & M & F & M  & F & M & F & M & \\
    \hline
20-34 & \makecell{16\\0.32} & \makecell{34\\1.42} & \makecell{40\\2.85} & \makecell{29\\9.81} & \makecell{16\\0.44} & \makecell{16\\1.41} & \makecell{29\\3.96} & \makecell{52\\7.88} & \makecell{28\\3.59} & \makecell{27\\2.94} & \makecell{52\\9.97} & \makecell{53\\10.59} &  \makecell{25\\3.71} & \makecell{30\\3.76} & \makecell{447\\62.65} \\
    \hline
35-49 & \makecell{21\\0.85} & \makecell{70\\2.94} & \makecell{37\\3.41} & \makecell{37\\8.07} & \makecell{24\\0.92} & \makecell{37\\1.87} & \makecell{48\\8.15} & \makecell{63\\12.24} &  \makecell{33\\4.32} & \makecell{44\\6.43} & \makecell{57\\11.28} &  \makecell{58\\11.22} & \makecell{36\\4.02} & \makecell{53\\7.47} & \makecell{618\\83.19} \\
    \hline
50-64 & \makecell{18\\0.62} & \makecell{52\\2.78} & \makecell{33\\4.91} & \makecell{34\\11.25} &  \makecell{27\\2.76} & \makecell{41\\2.33} & \makecell{45\\6.69} & \makecell{60\\13.98} &  \makecell{29\\5.96} & \makecell{47\\6.94} & \makecell{56\\11.73} &  \makecell{62\\12.02} & \makecell{26\\3.22} & \makecell{50\\5.53} & \makecell{580\\90.72} \\ 
    \hline
$\geq$65 & \makecell{21\\1.82} & \makecell{17\\1.11} & \makecell{29\\3.49} & \makecell{32\\9.25} & \makecell{18\\2.28} & \makecell{26\\4.55} & \makecell{21\\3.97} & \makecell{62\\15.21} & \makecell{31\\7.82} & \makecell{37\\6.13} & \makecell{48\\13.88} &  \makecell{59\\13.20} & \makecell{33\\5.14} & \makecell{30\\5.55} & \makecell{464\\93.4} \\
    \hhline{|=|=|=|=|=|=|=|=|=|=|=|=|=|=|=|=|}
Total & \makecell{76\\3.61} & \makecell{173\\8.25} & \makecell{139\\14.66} & \makecell{132\\38.38} & \makecell{85\\6.4} & \makecell{120\\10.16} & \makecell{143\\22.77} & \makecell{237\\49.31} & \makecell{121\\21.69} & \makecell{155\\22.44} & \makecell{213\\46.86} & \makecell{232\\47.03} & \makecell{120\\16.09} & \makecell{163\\22.31} & \makecell{2109\\329.96} \\
    \hline
\end{tabular}
 \end{adjustbox}
\caption{ 
Number of speakers (on top of the cell) and duration of recordings in hours (at the bottom of the cell) in each category of \textit{Age} (rows) by \textit{Period} and \textit{Gender} (columns) in the \textit{spINAch} corpus.}
\label{tab:corpus}
\end{table*}

\section{Data Analysis}
\label{sec:analysis}
This section presents a series of preliminary analyses of the corpus data, highlighting its versatility and showing how diachronic data reveal new insights into language description. 

\subsection{Does voice pitch change with time?} 
\label{sec:pitch}
Our voices are important parts of our personalities, as they index aspects of each individual, from their gender to their health \cite{Laver_1968, Eckert_2019, Podesva_Callier_2015}.
An important component of voice quality is related to its perceived pitch: a lower or higher voice being a central component of gender perception through vocal cues \cite{Leung_Oates_Chan_2018, Simpson_Weirich_2020}, but also to a series of interactive functions related to Ohala's \textit{Frequency Code} \cite{Ohala_1994}.
The construction of an individual's voice pitch is mediated by cultural components, notably those related to culturally variable representations of gender \cite{vanBezooijen_1995, Ohara_2001}.
As cultural values evolve with time, the role of women in society has undergone important changes since the end of the Second World War. 
As voice pitch is (negatively) related to social power in interaction \cite{Ohala_1994, Spencer-Oatey_1996, Goudbeek_Scherer_2010}, one may hypothesize that female voice pitch decreases over time.
A pitch decrease was claimed by some publications (\citealp[e.g.,][albeit without diachronic evidences]{Berg_Fuchs_Wirkner_Loeffler_Engel_Berger_2017}), but it is controversial in the literature (\citealp[e.g.,][]{Hollien_Hollien_DeJong_1997, Pemberton_McCormack_Russell_1998}).

The corpus presented here provides insight into potential changes in vocal characteristics for both genders among the subset of French personalities who are invited to radio and television shows.
Thanks to the acoustics measurements detailed in section~\ref{sec:acoustic}, we have a set of $f_o$ measurements, one for each oral vowel annotated as voiced (when both pitch detection algorithms returned coherent measures). 
There are 3,016,134 $f_o$ measurements, which are the median values observed on the voiced frames of each corresponding vowel.
The vowels come from 2,109 speakers, female or male, distributed uniquely across seven Periods as shown in Table~\ref{tab:corpus}.
The speakers' ages range from 20 to 95.

\subsubsection{Methods}
We tried to evaluate a potential evolution of $f_o$ across Periods, possibly linked to the speaker's gender, as a potentially different effect could be expected for female and male speakers, but controlling for changes linked to the speaker's age \cite{Berg_Fuchs_Wirkner_Loeffler_Engel_Berger_2017, Gisladottir_2023}.
Linear mixed-effect regression models were fitted (\citealp[following][]{Gries_2021, Crawley_2013}, and using R's \textit{lmer} library; \citealp{rcoreteam, bates2015}) to the median $f_o$ values of each vowel, expressed in semitones and standardized to avoid numerical problems.
The models took as predictors three fixed factors: the speaker's \textit{Age} (in years, centered around 50 years and divided by 30), their \textit{Gender} (two levels: ``F'' or ``M''), and the \textit{Period} of time corresponding to their recording (seven levels: 1955-56, 1965-66, 1975-76, 1985-86, 1995-96, 2005-06, 2015-16).
The models also controlled for variation associated with two random factors: the \textit{Speaker} (2,109 were considered) and the \textit{Vowel} category (12 levels). 
As vowel production is speaker-specific, the factor \textit{Vowel} was nested in \textit{Speaker}, itself nested in \textit{Gender} and in \textit{Period}.
The \textit{spINAch} corpus is cross-sectional, so each speaker belongs to a specific \textit{Gender} and a specific \textit{Period}.
\textit{Gender} was nested in \textit{Period} as gender representation in society may evolve across time.
Double interactions between each pair of the three random factors were also kept in the model, while the three-way interaction was discarded during a model simplification process \cite{Crawley_2013}, as not significant.
The model considered here follows equation~\ref{eq:modfo} (based on \textit{lmer}'s syntax), where $zf_o$ stands for the standardized $f_o,$ $A$ for \textit{Age}, $G$ for \textit{Gender}, $P$ for \textit{Period}, $S$ for \textit{Speaker}, and $V$ for \textit{Vowel} category.  
The square in equation~\ref{eq:modfo} encodes the two-way interactions among the $A$, $G$, and $P$ factors.
The ANOVA table for this model is presented in Table~\ref{tab:anovafo}.
\begin{equation} \label{eq:modfo}
zf_{o} \sim  (A + G + P)^2 + (1|P/G/S/V)
\end{equation}

\begin{table}[!ht]
\begin{center}
\begin{tabularx}{0.868\columnwidth}{lrrl}
      \toprule
        \textbf{Factors} & $\chi^2$ & \textbf{Df} & $Pr(>\chi^2)$ \\
      \midrule
$A$   &   6.9508 & 1 & 0.0083780 **  \\
$G$   &  14.9094 & 1 & 0.0001128 *** \\
$P$   &   0.0246 & 6 & 0.9999997     \\
$A\mathord{:}G$ &  57.6536 & 1 & 3.126e-14 *** \\
$A\mathord{:}P$ &  22.8561 & 6 & 0.0008461 *** \\
$G\mathord{:}P$ &   0.0593 & 6 & 0.9999958     \\
      \bottomrule
\end{tabularx}
\caption{ 
Analysis of Deviance Table (Type II Wald $\chi^2$ tests) obtained for the model fitted to the $f_o$ measurements; the $A$, $G$, and $P$ factors correspond to the \textit{Age}, \textit{Gender}, and \textit{Period} factors described in the text.}
\label{tab:anovafo}
\end{center}
\end{table}
\subsubsection{Results}
\begin{figure}
    \centering
    \includegraphics[width=\linewidth]{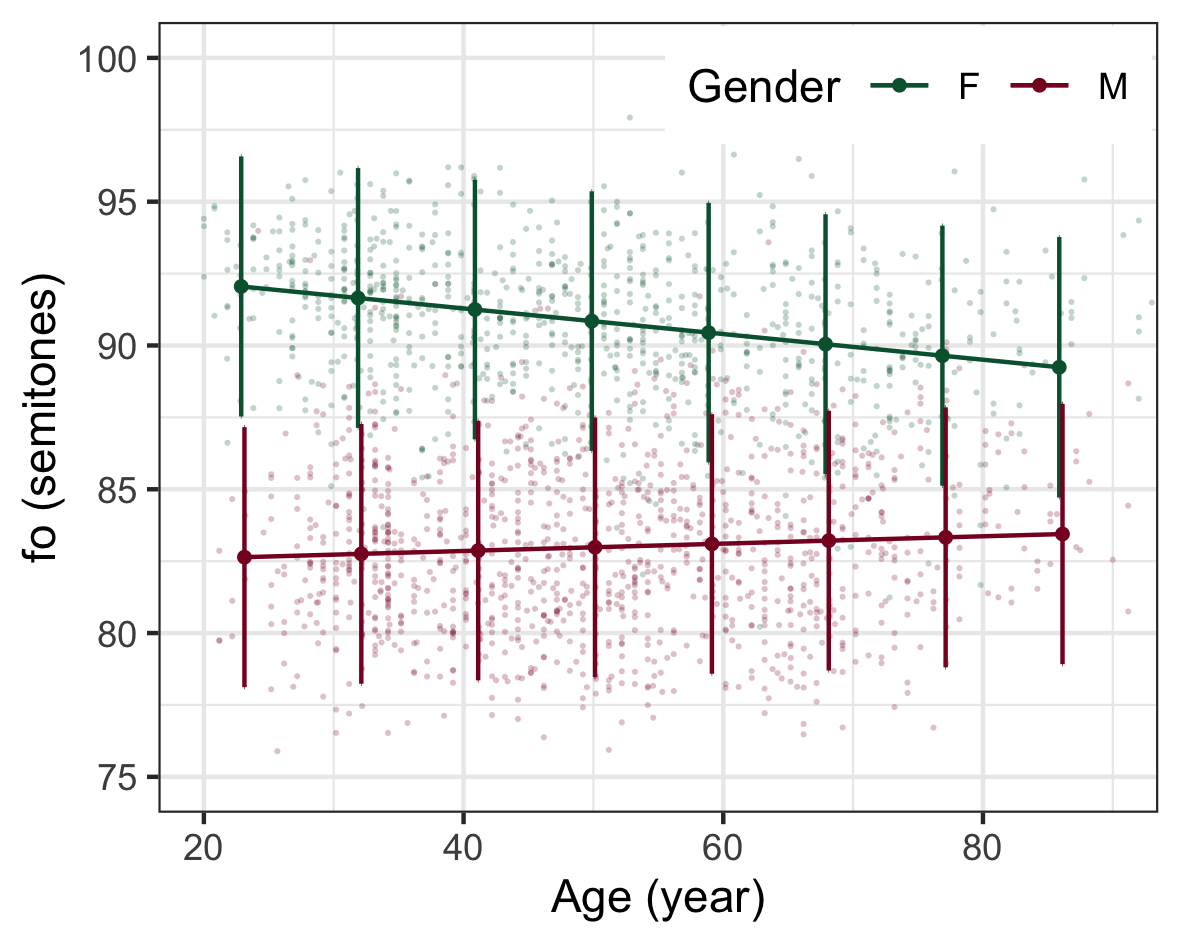}
    \caption{
    $f_o$ values (y-axis) estimated by the model for Age (x-axis) across Genders (colors), plotted over points that represent each speaker's mean $f_o$.}
    \label{fig:foAgeGender}
\end{figure}

Results showed that the speaker's \textit{Gender} has an (expected) major effect on $f_o$ values, with the mean $f_o$ difference of 7.7 semitones across genders, while, as main factors, \textit{Age} showed only a limited effect, and \textit{Period} had none.
\textit{Age} is nonetheless, and as described in the literature, fundamental to explain $f_o$ changes, but conditioning on the speaker's \textit{Gender} (cf. the $A\mathord{:}G$ line).
Female voices have their pitch lowered with \textit{Age} (a mean lowering of about 2.8 semitones in 60 years), while the reverse tendency is observed for male voices (a more modest rise of about 0.8 semitones in 60 years; see Figure~\ref{fig:foAgeGender}).

As for \textit{Period}, this factor has a limited impact on $f_o$ values. 
Its effect as a main factor is not significant, but it exhibits a significant interaction with \textit{Age}.
This interaction (not plotted for space reasons) is linked with the evolution of $f_o$ across speakers of different \textit{Ages} at a given \textit{Period}. 
This effect (controlled for Gender, as discussed above) consists of an increase of $f_o$ with age for the 1955-56 and 1965-66 \textit{Periods}, while the tendency is reversed starting with the 1985-86 \textit{Period}, when pitch tends to diminish with age.
This means that across older \textit{Periods}, as people get older, they tend to increase their pitch, controlling for their gender tendencies, whereas in newer periods, older individuals tend to decrease their pitch.
The effect size is comparatively limited with respect to the \textit{Age:Gender} interaction, but is still interesting, as potentially linked to varying social conditions in the population.
A possible explanation may relate to improved life expectancy in more recent periods, linked to better health conditions.

\subsection{Evolution of French vocalic system}
\label{sec:vowels}
One documented change in Parisian French during the twentieth century was an evolution of its vocalic system. 
Among other variations, the opposition between \textipa{/a/} and \textipa{/A/} is not productive in the main variant of French spoken in France (\citealp[for an overview of French diachrony, see][]{Abeille_Godard_2021}).
This vowel shift has already been observed on a series of speech corpora from different periods over a century \cite{Cecelewski_Gendrot_Adda-Decker_Boula_2024}, but on a dataset featuring only male speakers (because of the difficulty in finding female speakers in the archives).
We focus here on these two vowels (\textipa{/a/} an \textipa{/A/}), and not on the complete vocalic system of French, in order to give a simple example of diachronic changes and the importance of controlling for gender -- one of the key features of the \textit{spINAch} corpus.

\subsubsection{Methods}
These two vowels were annotated by the $MFA$ algorithm, which distinguishes between the two possible variants in its phonetic dictionary based on its acoustic model for French.
In the corpus, we get a sample of 
623,003 and 28,202 occurrences of \textipa{/a/} and \textipa{/A/} respectively, which first shows that they are clearly used unequally in French.
We fit two linear mixed-effect regression models, one on each of the first two formants ($F_1$ and $F_2$) expressed in bark and standardized. 
For $F_1$, after a simplification procedure, the model that was kept to describe the variation of this formant, controlling for variations linked to individual \textit{Speaker} ($S$), \textit{Gender} ($G$), \textit{Period} ($P$), \textit{Age} ($A$), and \textit{Vowel} category ($V$) corresponds to the Equation~\ref{eq:modF1}.
\begin{equation} \label{eq:modF1}
zF_{1} \sim  (A + G + P + V)^2 + A\mathord{:}P\mathord{:}V + (1|P/G/S/V)
\end{equation}
For $F_2$, a complex interaction between the speaker's age and the period was found. 
Following \cite{Stuart-Smith_2006}, we thus estimated each speaker's birth date, and evaluated the same model but using \textit{apparent time} in place of \textit{Period} and \textit{Age}. 
After simplification, the model corresponds to Equation~\ref{eq:modF2}, where $AT$ refers to the \textit{Apparent Time} when a given speaker starts learning their phonological system.
\begin{equation} \label{eq:modF2}
zF_{2} \sim  AT * (G + V) + (1|G/S/V)
\end{equation}

\subsubsection{Results}
The model fitted to the first formant (see Table~\ref{tab:anova_f1} in Appendix D) expects effects for the speaker's gender ($\chi_{(1)}^2=18.9, p<1.0\mathrm{e}{-4}$) and for the vowel category ($\chi_{(1)}^2=2618.8, p<1.0\mathrm{e}{-5}$). It also shows interactions between the Gender with the vowel category ($\chi_{(1)}^2=30.6, p<1.0\mathrm{e}{-5}$) and between the time Period with the vowel category ($\chi_{(6)}^2=15.9, p<0.05$) and a triple interaction between Period with vowel and the speaker's age ($\chi_{(6)}^2=13.8, p<0.05$).
The largest diachronic changes along $F_1$ are dependent on the vowel category, and consist of a $F_1$ decrease from 1955 to 1985, but the differences across the two vowel categories were mostly kept unchanged over time.
An increase of the first formant is known to follow the jaw aperture \cite{Erickson_2002}. 
Apart from $F_1$ obvious relation to vowel aperture, increased jaw aperture is also related to vocal effort (\citealp[e.g.,][]{Rilliard_dAlessandro_Evrard_2018}).
The observed diachronic changes in vocalic characteristics along this dimension may thus be linked to an evolution in recording practices in media outlets over this period, with a more declamatory style and microphones placed farther from the mouth in more ancient recordings (\citealp[see e.g.][]{BoulaRilliard_Allauzen_2012, devauchelle24}). 
In terms of the two vocalic categories, the differences across the \textipa{/a/} and \textipa{/A/} along $F_1$ do not change with time: it seems the vowels tagged as\textipa{/A/} had always larger jaw openings.

\begin{figure}
    \centering
    \includegraphics[width=\linewidth]{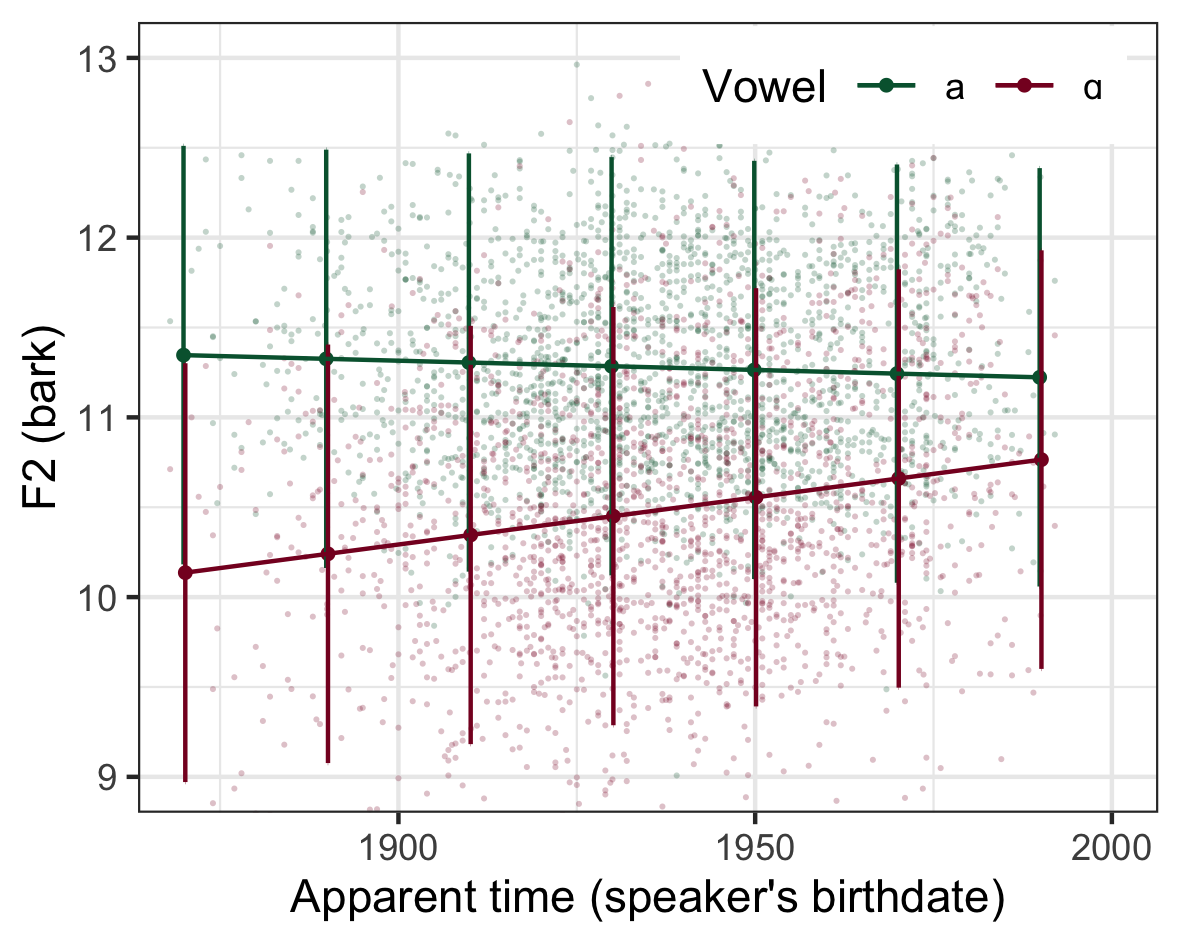}
    \caption{
    $F_2$ values (y-axis) estimated by the model for Apparent Time (x-axis) for both vowel categories (\textipa{/a/} and \textipa{/A/}; colors), plotted over points that represent each speaker's mean $F_2$.}
    \label{fig:f2at}
\end{figure}

For the model fitted on the second formant (see Table~\ref{tab:anova_f2} in Appendix D), we observed an expected effect of vowel category ($\chi_{(1)}^2=5377.1, p<1.0\mathrm{e}{-4}$) as both phones differ along the antero-posterior dimension.
There is also an interaction between the Apparent Time and the Vowel category ($\chi_{(1)}^2=204.4, p<1.0\mathrm{e}{-4}$) that is represented in Figure~\ref{fig:f2at}. 
It shows that the two vowels' $F_2$ values converge over the twentieth century, so they are statistically comparable for people born around the middle of the century.
We also observe a significant interaction between Apparent Time and the speaker's gender ($\chi_{(1)}^2=11.6, p<1.0\mathrm{e}{-3}$), that is independent of vowel category.
Because of a longer vocal tract, males tend to have lower $F_2$ values than females, but male speakers born more recently tend to amplify this difference, once controlled for vowel: they display a lowering $F_2$ trend over Apparent Time that is not observed for female speakers (whose mean $F_2$ values are almost flat with time). 

This continued lowering of $F_2$ for these two vowel categories along the \textsc{xx}\textsuperscript{th}~century is visible in the analysis proposed by \citealt{Cecelewski_Gendrot_Adda-Decker_Boula_2024} -- who worked on male-only datasets. 
The interpretation of a gender-specific tendency supported by our dataset may thus differ from theirs.
Lowering $F_2$ is a correlate of a posteriorized articulation, an articulatory setting that tends to lower a voice's pitch, which may be a marker of masculinity \cite{vanBezooijen_1995} -- while the contrary, an anteriorization strategy, was linked by other works to a phonostyle of seduction used by French females \cite{Leon_1993, Rilliard_dAlessandro_Evrard_2018}.
So, during the first half of the century, we observe in our dataset a convergence of the two phonetic categories, with the/\textipa{A}/ vowel progressively anteriorized until it mixes with /\textipa{a}/; this evolution was not gender-dependent.
Once the vocalic system has only one category of open vowel, it has more space for sociophonetic variations -- and during the second half of the century, Parisian French-speaking males appear to use this available space to reinforce their voice's masculinity by posteriorizing their articulation of the now single open vowel. 
Such a gender-specific tendency was not observed in our diachronic data for female speakers (who kept a more anterior articulation, compared to males). 
Working on a subset of the \textit{spINAch} corpus, \citet{elie24b} had already observed gender-exaggerating articulatory tendencies in female and male speakers, who are using lips and larynx positions to respectively shorten or lengthen their vocal tract. 
Meanwhile, in \citet{elie24b}'s work, the tendency did not evolve diachronically.
Here, the extra vocalic space left by the fusion of /\textipa{a}/ and /\textipa{A}/ appears to have been used for indexical means by male speakers.

\section{Conclusion}
In this paper, we present the \textit{spINAch} corpus, featuring more than 320 hours of speech extracted from radio and television broadcasts over a 60-year period, from speakers selected to balance gender and age distributions.
This dataset comprises more than two thousand speakers, represented by speech segments of varying duration, with a median speaker duration of over six minutes (390~s). 
The speaker's birth dates range from 1870 to 1990.
The corpus, which is made available for research purposes\footnote{Available at \url{https://www.ina.fr/institut-national-audiovisuel/research/dataset-project\#spINAch}}, contains the audio samples, their automatic transcription, and the phonetic alignment.
The speakers' identities are not disclosed, with only non-identifying demographic information retained (age at the time of recording, with a five-year precision, and gender).
All audio segments of a given speaker resulting from the diarization process have been assigned a random ID to make it difficult to reconstruct the speaker's argument for potential authorship purposes.
The acoustic analyses ($f_o$ and formants) presented in this paper focused on oral vowels. More than 3 million vowels have been analyzed, and these measurements are presented in a separate file to allow interested researchers to study this aspect of the corpus directly\footnote{Available at \url{https://doi.org/10.5281/zenodo.18714702}}.

The two rapid analyses of this dataset presented in section~\ref{sec:analysis}, beyond their specific findings, showed that the \textit{spINAch} corpus contains reliable data, as we were able to reproduce several known characteristics of $f_o$ changes with age or vocalic variation in French. 
These data enable a variety of phonetic investigations of changes in French as it is spoken in France's national media outlets over this 60-year period. 
We hope our efforts to gather this dataset will enable the community to conduct more research on the diachrony of the French language.

\section{Acknowledgments}
This work was partly funded by ANR "Gender Equality Monitor" (GEM) grant ANR-19-CE38-0012. 
We are especially grateful to Pascal Flard at INA for his valuable assistance in recovering part of the archives needed for this work.

\section{Bibliographical References}\label{sec:reference}

\bibliographystyle{lrec2026-natbib}
\bibliography{lrec2026-example}

\begin{thebibliography}{4}
\expandafter\ifx\csname natexlab\endcsname\relax\def\natexlab#1{#1}\fi

\bibitem[{Ardila et~al.(2020)Ardila, Branson, Davis, Kohler, Meyer, Henretty,
  Morais, Saunders, Tyers, and Weber}]{ardila-etal-2020-common}
Ardila, Rosana and Branson, Megan and Davis, Kelly and Kohler, Michael and
  Meyer, Josh and Henretty, Michael and Morais, Reuben and Saunders, Lindsay
  and Tyers, Francis and Weber, Gregor. 2020.
\newblock \href {https://aclanthology.org/2020.lrec-1.520/} {\emph{Common
  Voice: A Massively-Multilingual Speech Corpus}}.
\newblock European Language Resources Association.

\bibitem[{Meléndez-Catalán et~al.(2019)Meléndez-Catalán, Molina, and
  Gómez}]{MelndezCataln2019}
Meléndez-Catalán, Blai and Molina, Emilio and Gómez, Emilia. 2019.
\newblock \href {https://doi.org/10.5334/tismir.29} {\emph{Open Broadcast Media
  Audio from TV: A Dataset of TV Broadcast Audio with Relative Music Loudness
  Annotations}}.
\newblock Ubiquity Press, Ltd.

\bibitem[{Nagrani et~al.(2017)Nagrani, Chung, and
  Zisserman}]{Nagrani_Chung_Zisserman_2017}
Nagrani, Arsha and Chung, Joon Son and Zisserman, Andrew. 2017.
\newblock \href {https://doi.org/10.21437/Interspeech.2017-950}
  {\emph{VoxCeleb: A Large-Scale Speaker Identification Dataset}}.
\newblock ISCA.

\bibitem[{Seyerlehner et~al.(2007)Seyerlehner, Pohle, Schedl, and
  Widmer}]{seyerlehner2007automatic}
Seyerlehner, Klaus and Pohle, Tim and Schedl, Markus and Widmer, Gerhard. 2007.
\newblock \emph{Automatic music detection in television productions}.
\newblock SCRIME/LaBRI Bordeaux.

\end{thebibliography}


\begin{thebibliography}{59}
\expandafter\ifx\csname natexlab\endcsname\relax\def\natexlab#1{#1}\fi

\bibitem[{Abeillé and Godard(2021)}]{Abeille_Godard_2021}
Anne Abeillé and Danièle Godard. 2021.
\newblock \emph{La grande grammaire du français}.
\newblock Actes Sud Imprimerie nationale éditions, Arles [Paris].

\bibitem[{Ballier and Méli(2024)}]{Ballier_Meli_2024}
Nicolas Ballier and Adrien Méli. 2024.
\newblock \href {https://doi.org/10.3384/ecp211002} {Investigating acoustic
  correlates of whisper scoring for l2 speech using forced alignment with the
  italian component of the isle corpus}.
\newblock page 20–32.

\bibitem[{Barras et~al.(2002)Barras, Allauzen, Lamel, and
  Gauvain}]{Barras_Allauzen_Lamel_Gauvain_2002}
Claude Barras, Alexandre Allauzen, Lori Lamel, and Jean-Luc Gauvain. 2002.
\newblock \href {https://doi.org/10.1109/ICASSP.2002.5743642} {Transcribing
  audio-video archives}.
\newblock In \emph{IEEE International Conference on Acoustics Speech and Signal
  Processing}, pages I--13--I–16, Orlando, FL, USA. IEEE.

\bibitem[{Bates et~al.(2015)Bates, M{\"a}chler, Bolker, and Walker}]{bates2015}
Douglas Bates, Martin M{\"a}chler, Ben Bolker, and Steve Walker. 2015.
\newblock \href {https://doi.org/10.18637/jss.v067.i01} {Fitting linear
  mixed-effects models using {lme4}}.
\newblock \emph{Journal of Statistical Software}, 67(1):1--48.

\bibitem[{Berg et~al.(2017)Berg, Fuchs, Wirkner, Loeffler, Engel, and
  Berger}]{Berg_Fuchs_Wirkner_Loeffler_Engel_Berger_2017}
Martin Berg, Michael Fuchs, Kerstin Wirkner, Markus Loeffler, Christoph Engel,
  and Thomas Berger. 2017.
\newblock \href {https://doi.org/10.1016/j.jvoice.2016.06.001} {The speaking
  voice in the general population: Normative data and associations to
  sociodemographic and lifestyle factors}.
\newblock \emph{Journal of Voice}, 31(2):257.e13--257.e24.

\bibitem[{Boersma(1993)}]{Boersma_1993}
Paul Boersma. 1993.
\newblock Accurate short-term analysis of the fundamental frequency and the
  harmonics-to-noise ratio of a sampled sound.
\newblock \emph{Proceedings of the institute of phonetic sciences},
  17:97–110.

\bibitem[{Boersma and Weenink(2025)}]{Boersma_Weenink_2025}
Paul Boersma and David Weenink. 2025.
\newblock \href {http://www.praat.org/} {Praat: doing phonetics by computer
  [computer program]. version 6.4.45}.

\bibitem[{Boula~de Mareüil et~al.(2012)Boula~de Mareüil, Rilliard, and
  Allauzen}]{BoulaRilliard_Allauzen_2012}
Philippe Boula~de Mareüil, Albert Rilliard, and Alexandre Allauzen. 2012.
\newblock \href {https://doi.org/10.1177/0023830911417799} {A diachronic study
  of initial stress and other prosodic features in the french news announcer
  style: Corpus-based measurements and perceptual experiments}.
\newblock \emph{Language and Speech}, 55(2):263–293.

\bibitem[{Bredin(2023)}]{bredin23_interspeech}
Hervé Bredin. 2023.
\newblock \href {https://doi.org/10.21437/Interspeech.2023-105} {pyannote.audio
  2.1 speaker diarization pipeline: principle, benchmark, and recipe}.
\newblock In \emph{Interspeech 2023}, pages 1983--1987.

\bibitem[{Christodoulidou et~al.(2025)Christodoulidou, Tanner, Stuart-Smith,
  McAuliffe, Murali, Smith, Taylor, Cleland, and
  Kuschmann}]{christodoulidou25_interspeech}
Polychronia Christodoulidou, James Tanner, Jane Stuart-Smith, Michael
  McAuliffe, Mridhula Murali, Amy Smith, Lauren Taylor, Joanne Cleland, and
  Anja Kuschmann. 2025.
\newblock \href {https://doi.org/10.21437/Interspeech.2025-1030} {{A
  semi-automatic pipeline for transcribing and segmenting child speech}}.
\newblock In \emph{{Interspeech 2025}}, pages 4278--4282.

\bibitem[{Coats(2025)}]{Coats_2025}
Steven Coats. 2025.
\newblock \href {https://doi.org/10.1515/9783111434018-011} {\emph{An automatic
  pipeline for processing streamed content: New horizons for corpus linguistics
  and phonetics}}, page 257–274. De Gruyter.

\bibitem[{Coulomb-Gully(2011)}]{CoulombGully_2011}
Marlène Coulomb-Gully. 2011.
\newblock Genre et médias~: vers un état des lieux.
\newblock \emph{Sciences de la société}, 83:3--13.

\bibitem[{Crawley(2013)}]{Crawley_2013}
Michael~J. Crawley. 2013.
\newblock \emph{The R book}, second edition edition.
\newblock Wiley, Chichester, West Sussex, UK.

\bibitem[{Cęcelewski et~al.(2024)Cęcelewski, Gendrot, Adda-Decker, and
  Boula~de Mareüil}]{Cecelewski_Gendrot_Adda-Decker_Boula_2024}
Juliusz Cęcelewski, Cédric Gendrot, Martine Adda-Decker, and Philippe
  Boula~de Mareüil. 2024.
\newblock \href {https://aclanthology.org/2024.jeptalnrecital-jep.8/} {Étude
  en temps réel de la fusion des /a/ \textasciitilde /\textipa{A}/ en
  français depuis 1925}.
\newblock In \emph{Actes des 35èmes Journées d’Études sur la Parole}, page
  71–81, Toulouse, France. ATALA and AFPC.

\bibitem[{Devauchelle et~al.(2024)Devauchelle, Rilliard, Doukhan, and
  Yang}]{devauchelle24}
Simon Devauchelle, Albert Rilliard, David Doukhan, and Lucas~Ondel Yang. 2024.
\newblock \href {https://doi.org/10.17469/O2112AISV000004} {{Variation of
  Perceived Voice Pitch Across Time Periods, Gender, and Age in French Media
  Archives}}.
\newblock In Valentina~De Iacovo, Bianca Maria~De Paolis, and Daniela Mereu,
  editors, \emph{{The voice in the media and new technologies}}, volume~12 of
  \emph{Studi Associazione Italiana Scienze della Voce}, pages 47--71.
  {Officinaventuno}.

\bibitem[{Doukhan et~al.(2018{\natexlab{a}})Doukhan, Carrive, Vallet, Larcher,
  and Meignier}]{Doukhan_2018}
David Doukhan, Jean Carrive, Félicien Vallet, Anthony Larcher, and Sylvain
  Meignier. 2018{\natexlab{a}}.
\newblock An open-source speaker gender detection framework for monitoring
  gender equality.
\newblock In \emph{Acoustics Speech and Signal Processing (ICASSP), 2018 IEEE
  International Conference on}. IEEE.

\bibitem[{Doukhan et~al.(2018{\natexlab{b}})Doukhan, Poels, Rezgui, and
  Carrive}]{Doukhan_2018B}
David Doukhan, Géraldine Poels, Zohra Rezgui, and Jean Carrive.
  2018{\natexlab{b}}.
\newblock \href {https://doi.org/http://dx.doi.org/10.25969/mediarep/1475}
  {Describing gender equality in french audiovisual streams with a deep
  learning approach}.
\newblock \emph{VIEW Journal of European Television History and Culture},
  7(14):103--122.

\bibitem[{Eckert(2019)}]{Eckert_2019}
Penelope Eckert. 2019.
\newblock \href {https://doi.org/10.1353/lan.2019.0072} {The limits of meaning:
  Social indexicality, variation, and the cline of interiority}.
\newblock \emph{Language}, 95(4):751–776.

\bibitem[{Elie et~al.(2024)Elie, Doukhan, Uro, Ondel-Yang, Rilliard, and
  Devauchelle}]{elie24b}
Benjamin Elie, David Doukhan, Rémi Uro, Lucas Ondel-Yang, Albert Rilliard, and
  Simon Devauchelle. 2024.
\newblock \href {https://doi.org/10.21437/Interspeech.2024-1177} {{Articulatory
  Configurations across Genders and Periods in French Radio and TV archives}}.
\newblock In \emph{{Interspeech 2024}}, pages 3085--3089.

\bibitem[{Erickson(2002)}]{Erickson_2002}
Donna Erickson. 2002.
\newblock \href {https://doi.org/10.1159/000066067} {Articulation of extreme
  formant patterns for emphasized vowels}.
\newblock \emph{Phonetica}, 59(2–3):134–149.

\bibitem[{Escudero et~al.(2009)Escudero, Boersma, Rauber, and
  Bion}]{Escudero_Boersma_Rauber_Bion_2009}
Paola Escudero, Paul Boersma, Andréia~Schurt Rauber, and Ricardo A.~H. Bion.
  2009.
\newblock \href {https://doi.org/10.1121/1.3180321} {A cross-dialect acoustic
  description of vowels: Brazilian and european portuguese}.
\newblock \emph{The Journal of the Acoustical Society of America},
  126(3):1379–1393.

\bibitem[{Gisladottir et~al.(2023)Gisladottir, Helgason, Halldorsson, Helgason,
  Borsky, Chien, Gudnason, Gudjonsson, Moisik, Dediu, Thorleifsson, Tragante,
  Bustamante, Jonsdottir, Stefansdottir, Rutsdottir, Magnusson, Hardarson,
  Ferkingstad, Halldorsson, Rognvaldsson, Skuladottir, Ivarsdottir, Norddahl,
  Thorgeirsson, Jonsdottir, Ulfarsson, Holm, Stefansson, Thorsteinsdottir,
  Gudbjartsson, Sulem, and Stefansson}]{Gisladottir_2023}
Rosa~S. Gisladottir, Agnar Helgason, Bjarni~V. Halldorsson, Hannes Helgason,
  Michal Borsky, Yu-Ren Chien, Jon Gudnason, Sigurjon~A. Gudjonsson, Scott
  Moisik, Dan Dediu, Gudmar Thorleifsson, Vinicius Tragante, Mariana
  Bustamante, Gudrun~A. Jonsdottir, Lilja Stefansdottir, Gudrun Rutsdottir,
  Sigurdur~H. Magnusson, Marteinn Hardarson, Egil Ferkingstad, Gisli~H.
  Halldorsson, Solvi Rognvaldsson, Astros Skuladottir, Erna~V. Ivarsdottir,
  Gudmundur Norddahl, Gudmundur Thorgeirsson, Ingileif Jonsdottir, Magnus~O.
  Ulfarsson, Hilma Holm, Hreinn Stefansson, Unnur Thorsteinsdottir, Daniel~F.
  Gudbjartsson, Patrick Sulem, and Kari Stefansson. 2023.
\newblock \href {https://doi.org/10.1126/sciadv.abq2969} {Sequence variants
  affecting voice pitch in humans}.
\newblock \emph{Science Advances}, 9(23):eabq2969.

\bibitem[{Goudbeek and Scherer(2010)}]{Goudbeek_Scherer_2010}
Martijn Goudbeek and Klaus Scherer. 2010.
\newblock \href {https://doi.org/10.1121/1.3466853} {Beyond arousal: Valence
  and potency/control cues in the vocal expression of emotion}.
\newblock \emph{The Journal of the Acoustical Society of America},
  128(3):1322–1336.

\bibitem[{Gries(2021)}]{Gries_2021}
Stefan~Thomas Gries. 2021.
\newblock \emph{Statistics for linguistics with R: a practical introduction},
  3rd revised edition edition.
\newblock De Gruyter Mouton textbook. de Gruyter Mouton, Berlin Boston.

\bibitem[{Harrington et~al.(2000)Harrington, Palethorpe, and
  Watson}]{Harrington_Palethorpe_Watson_2000}
Jonathan Harrington, Sallyanne Palethorpe, and Catherine Watson. 2000.
\newblock \href {https://doi.org/10.1017/S0025100300006666} {Monophthongal
  vowel changes in received pronunciation: an acoustic analysis of the
  queen’s christmas broadcasts}.
\newblock \emph{Journal of the International Phonetic Association},
  30(1–2):63–78.

\bibitem[{Hollien et~al.(1994)Hollien, Green, and
  Massey}]{Hollien_Green_Massey_1994}
Harry Hollien, Rachel Green, and Karen Massey. 1994.
\newblock \href {https://doi.org/10.1121/1.411275} {Longitudinal research on
  adolescent voice change in males}.
\newblock \emph{The Journal of the Acoustical Society of America},
  96(5):2646–2654.

\bibitem[{Hollien et~al.(1997)Hollien, Hollien, and
  De~Jong}]{Hollien_Hollien_DeJong_1997}
Harry Hollien, Patricia~A. Hollien, and Gea De~Jong. 1997.
\newblock \href {https://doi.org/10.1121/1.420353} {Effects of three parameters
  on speaking fundamental frequency}.
\newblock \emph{The Journal of the Acoustical Society of America},
  102(5):2984–2992.

\bibitem[{Kuznetsova et~al.(2017)Kuznetsova, Brockhoff, and
  Christensen}]{Kuznetsova_2017}
Alexandra Kuznetsova, Per~B. Brockhoff, and Rune H.~B. Christensen. 2017.
\newblock \href {https://doi.org/10.18637/jss.v082.i13} {{lmerTest} package:
  Tests in linear mixed effects models}.
\newblock \emph{Journal of Statistical Software}, 82(13):1--26.

\bibitem[{Laver(1968)}]{Laver_1968}
John D.~M. Laver. 1968.
\newblock \href {https://doi.org/10.3109/13682826809011440} {Voice quality and
  indexical information}.
\newblock \emph{British Journal of Disorders of Communication}, 3(1):43–54.

\bibitem[{Leung et~al.(2018)Leung, Oates, and Chan}]{Leung_Oates_Chan_2018}
Yeptain Leung, Jennifer Oates, and Siew~Pang Chan. 2018.
\newblock \href {https://doi.org/10.1044/2017_JSLHR-S-17-0067} {Voice,
  articulation, and prosody contribute to listener perceptions of speaker
  gender: A systematic review and meta-analysis}.
\newblock \emph{Journal of Speech, Language, and Hearing Research},
  61(2):266–297.

\bibitem[{Li et~al.(2022)Li, Yuan, Zhang, MA, Lin, Chen, Ragni, Yin, Hu, He,
  Benetos, Gyenge, Liu, and Fu}]{li2022mapmusic2vecsimpleeffectivebaseline}
Y.~Li, R.~Yuan, G.~Zhang, Y.~MA, C.~Lin, X.~Chen, A.~Ragni, H.~Yin, Z.~Hu,
  H.~He, E.~Benetos, N.~Gyenge, R.~Liu, and J.~Fu. 2022.
\newblock \href {https://ismir2022program.ismir.net/lbd\%5f410.html} {Lv-49:
  Map-music2vec: A simple and effective baseline for self-supervised music
  audio representation learning}.
\newblock In \emph{23rd International Society for Music Information Retrieval
  Conference (ISMIR 2022)}.

\bibitem[{Léon(1993)}]{Leon_1993}
Pierre Léon. 1993.
\newblock \emph{Précis de phonostylistique. Parole et expressivité}.
\newblock Nathan Université, Paris.

\bibitem[{McAuliffe et~al.(2017)McAuliffe, Socolof, Mihuc, Wagner, and
  Sonderegger}]{McAuliffe_Socolof_Mihuc_Wagner_Sonderegger_2017}
Michael McAuliffe, Michaela Socolof, Sarah Mihuc, Michael Wagner, and Morgan
  Sonderegger. 2017.
\newblock \href {https://doi.org/10.21437/Interspeech.2017-1386} {Montreal
  forced aligner: Trainable text-speech alignment using kaldi}.
\newblock In \emph{Interspeech 2017}, page 498–502. ISCA.

\bibitem[{Meignier and Merlin(2010)}]{Meignier_2010}
Sylvain Meignier and Teva Merlin. 2010.
\newblock Lium spkdiarization: an open source toolkit for diarization.
\newblock In \emph{CMU SPUD Workshop}.

\bibitem[{Mufwene(2007)}]{Mufwene_2007}
Salikoko~S Mufwene. 2007.
\newblock Population movements and contacts in language evolution.
\newblock \emph{Journal of language contact}, 1(1):63–92.

\bibitem[{Ménard et~al.(2009)Ménard, Dupont, Baum, and
  Aubin}]{Menard_Dupont_Baum_Aubin_2009}
Lucie Ménard, Sophie Dupont, Shari~R. Baum, and Jérôme Aubin. 2009.
\newblock \href {https://doi.org/10.1121/1.3158930} {Production and perception
  of french vowels by congenitally blind adults and sighted adults}.
\newblock \emph{The Journal of the Acoustical Society of America},
  126(3):1406–1414.

\bibitem[{Ohala(1994)}]{Ohala_1994}
John~J. Ohala. 1994.
\newblock \href {https://doi.org/10.1017/CBO9780511751806.022} {\emph{The
  frequency code underlies the sound-symbolic use of voice pitch}}, 1 edition,
  page 325–347. Cambridge University Press.

\bibitem[{Ohara(2001)}]{Ohara_2001}
Yumiko Ohara. 2001.
\newblock \href {https://doi.org/10.1515/9783110889406.231} {\emph{Finding
  one’s voice in Japanese: A study of the pitch levels of L2 users}}. DE
  GRUYTER MOUTON, Berlin, New York.

\bibitem[{Pelloin et~al.(2026)Pelloin, Bekkali, Dehak, and
  Doukhan}]{pelloin_lrec2026}
Valentin Pelloin, Lina Bekkali, Reda Dehak, and David Doukhan. 2026.
\newblock Data selection effects on self-supervised learning of audio
  representations for french audiovisual broadcasts.
\newblock In \emph{Fifteenth International Conference on Language Resources and
  Evaluation (LREC 2026)}, Palma, Mallorca, Spain. European Language Resources
  Association.

\bibitem[{Pemberton et~al.(1998)Pemberton, McCormack, and
  Russell}]{Pemberton_McCormack_Russell_1998}
Cecilia Pemberton, Paul McCormack, and Alison Russell. 1998.
\newblock \href {https://doi.org/10.1016/S0892-1997(98)80040-4} {Have women’s
  voices lowered across time? a cross sectional study of australian women’s
  voices}.
\newblock \emph{Journal of Voice}, 12(2):208–213.

\bibitem[{Podesva and Callier(2015)}]{Podesva_Callier_2015}
Robert~J. Podesva and Patrick Callier. 2015.
\newblock \href {https://doi.org/10.1017/S0267190514000270} {Voice quality and
  identity}.
\newblock \emph{Annual Review of Applied Linguistics}, 35:173–194.

\bibitem[{{R Core Team}(2024)}]{rcoreteam}
{R Core Team}. 2024.
\newblock \href {https://www.R-project.org/} {\emph{R: A Language and
  Environment for Statistical Computing}}.
\newblock R Foundation for Statistical Computing, Vienna, Austria.

\bibitem[{Radford et~al.(2022)Radford, Kim, Xu, Brockman, McLeavey, and
  Sutskever}]{Radford_2022}
Alec Radford, Jong~Wook Kim, Tao Xu, Greg Brockman, Christine McLeavey, and
  Ilya Sutskever. 2022.
\newblock \href {http://arxiv.org/abs/2212.04356} {Robust speech recognition
  via large-scale weak supervision}.

\bibitem[{Rilliard et~al.(2018)Rilliard, d’Alessandro, and
  Evrard}]{Rilliard_dAlessandro_Evrard_2018}
Albert Rilliard, Christophe d’Alessandro, and Marc Evrard. 2018.
\newblock \href {https://doi.org/10.1121/1.5018433} {Paradigmatic variation of
  vowels in expressive speech: Acoustic description and dimensional analysis}.
\newblock \emph{The Journal of the Acoustical Society of America},
  143(1):109–122.

\bibitem[{Riverin-Coutlée and
  Harrington(2022)}]{Riverin-Coutlee_Harrington_2022}
Josiane Riverin-Coutlée and Jonathan Harrington. 2022.
\newblock \href {https://doi.org/10.1515/lingvan-2021-0122} {Phonetic change
  over the career: a case study}.
\newblock \emph{Linguistics Vanguard}, 8(1):41–52.

\bibitem[{Simpson and Weirich(2020)}]{Simpson_Weirich_2020}
Adrian~P. Simpson and Melanie Weirich. 2020.
\newblock \href {https://doi.org/10.1093/acrefore/9780199384655.013.749}
  {\emph{Phonetic Correlates of Sex, Gender and Sexual Orientation}}. Oxford
  University Press.

\bibitem[{Sloetjes and Wittenburg(2008)}]{sloetjes2008annotation}
Han Sloetjes and Peter Wittenburg. 2008.
\newblock Annotation by category-elan and iso dcr.
\newblock In \emph{6th international Conference on Language Resources and
  Evaluation (LREC 2008)}.

\bibitem[{Spencer-Oatey(1996)}]{Spencer-Oatey_1996}
Helen Spencer-Oatey. 1996.
\newblock \href {https://doi.org/10.1016/0378-2166(95)00047-X} {Reconsidering
  power and distance}.
\newblock \emph{Journal of Pragmatics}, 26(1):1–24.

\bibitem[{Stuart-Smith(2006)}]{Stuart-Smith_2006}
Jane Stuart-Smith. 2006.
\newblock \href {https://doi.org/10.4324/9780203441497} {\emph{The Influence of
  the Media}}, 0 edition, page 140–148. Routledge.

\bibitem[{Stuart-Smith(2020)}]{Stuart-Smith_2020}
Jane Stuart-Smith. 2020.
\newblock \href {https://doi.org/10.1515/lingvan-2018-0064} {Changing
  perspectives on /s/ and gender over time in glasgow}.
\newblock \emph{Linguistics Vanguard}, 6(s1):20180064.

\bibitem[{Suire and Barkat-Defradas(2020)}]{Suire_Barkat-Defradas_2020}
Alexandre Suire and Melissa Barkat-Defradas. 2020.
\newblock Evolution of human pitch: Preliminary analyses in the french
  population using ina audiovisual archives of vox pops.
\newblock In \emph{2020 IASA-FIAT/IFTA Joint Conference}.

\bibitem[{Talkin(2015)}]{Talkin_2015}
David Talkin. 2015.
\newblock \href {https://github.com/google/REAPER} {Reaper: Robust epoch and
  pitch estimator}.

\bibitem[{Traunmüller(1990)}]{Traunmuller_1990}
Hartmut Traunmüller. 1990.
\newblock \href {https://doi.org/10.1121/1.399849} {Analytical expressions for
  the tonotopic sensory scale}.
\newblock \emph{The Journal of the Acoustical Society of America}, 88:97--100.

\bibitem[{Uro et~al.(2022)Uro, Doukhan, Rilliard, Larcher, Adgharouamane,
  Tahon, and Laurent}]{uro22}
R{\'e}mi Uro, David Doukhan, Albert Rilliard, Laetitia Larcher, Anissa-Claire
  Adgharouamane, Marie Tahon, and Antoine Laurent. 2022.
\newblock \href {https://aclanthology.org/2022.lrec-1.350/} {A semi-automatic
  approach to create large gender- and age-balanced speaker corpora: Usefulness
  of speaker diarization {\&} identification.}
\newblock In \emph{Proceedings of the Thirteenth Language Resources and
  Evaluation Conference}, pages 3271--3280, Marseille, France. European
  Language Resources Association.

\bibitem[{Vallet and Carrive(2014)}]{vallet14}
Félicien Vallet and Jean Carrive. 2014.
\newblock Quand l'horloge parlante a beaucoup à raconter sur l'évolution des
  techniques d'archivage audiovisuel.
\newblock In \emph{Journées d'étude sur la parole}.

\bibitem[{van Bezooijen(1995)}]{vanBezooijen_1995}
Reneé van Bezooijen. 1995.
\newblock \href {https://doi.org/10.1177/002383099503800303} {Sociocultural
  aspects of pitch differences between japanese and dutch women}.
\newblock \emph{Language and Speech}, 38(3):253–265.

\bibitem[{Vaysse et~al.(2022)Vaysse, Astésano, and
  Farinas}]{Vaysse_Astesano_Farinas_2022}
Robin Vaysse, Corine Astésano, and Jérôme Farinas. 2022.
\newblock \href {https://doi.org/10.1121/10.0015143} {Performance analysis of
  various fundamental frequency estimation algorithms in the context of
  pathological speech}.
\newblock \emph{The Journal of the Acoustical Society of America},
  152(5):3091–3101.

\bibitem[{Vigouroux(2015)}]{Vigouroux_2015}
Cécile~B. Vigouroux. 2015.
\newblock \href {https://doi.org/10.1017/S0047404515000068} {Genre,
  heteroglossic performances, and new identity: Stand-up comedy in modern
  french society}.
\newblock \emph{Language in Society}, 44(2):243–272.

\bibitem[{Zou et~al.(2012)Zou, Wang, and He}]{Zou_Wang_He_2012}
Yu~Zou, Yan Wang, and Wei He. 2012.
\newblock \href {https://doi.org/10.1109/ISCSLP.2012.6423498} {Diachronic
  contrastive analysis on read speech in broadcast news: Evidence from pitch
  and duration}.
\newblock In \emph{2012 8th International Symposium on Chinese Spoken Language
  Processing}, page 291–295, Kowloon Tong, China. IEEE.

\end{thebibliography}

\section{Language Resource References}
\label{lr:ref}
\bibliographystylelanguageresource{lrec2026-natbib}
\bibliographylanguageresource{languageresource}

\section{Supplementary Materials}

\subsection{A : ANOVA Tables} 
\label{sec:Appendix_A}

ANOVA table for models presented in section~\ref{sec:evalex}, fitted for the first three formants (expressed in Bark and standardized): Table~\ref{tab:anova_fixed} presents the results for the three fixed factors (\textit{Method}, \textit{Gender}, and \textit{Vowel}), while Table~\ref{tab:anova_random} presents the effect of the random factors, obtained using single-term deletion with the \textit{lmerTest} library \cite{Kuznetsova_2017}.

\begin{table*}
    \begin{center}
    \begin{adjustbox}{width=0.43\textwidth}
    \begin{tabular}{llrrl}
          \toprule
            \textbf{Model} & \textbf{Factors} & $\chi^2$ & \textbf{Df} & $Pr(>\chi^2)$ \\
          \midrule
    $F_1$ & Method   &     0.08   &  1 & 0.772     \\
          & Gender   &    17.37   &  1 & 0.002 **  \\
          & Vowel    &   202660   & 11 & 0.000 *** \\
          \midrule
    $F_2$ & Method   &    0.96    &  1 & 0.327     \\
          & Gender   &   68.24    &  1 & 0.000 *** \\
          & Vowel    &  287610    & 11 & 0.000 *** \\
          \midrule
    $F_3$ & Method   &    0.61   &  1 & 0.433     \\
          & Gender   &   49.72   &  1 & 0.000 *** \\
          & Vowel    &   69210   & 11 & 0.000 *** \\
          \bottomrule
    \end{tabular}
    \end{adjustbox}
    \caption{ 
    \centering Analysis of Deviance Table (Type II Wald $\chi^2$ tests) obtained for the fixed factors of the models fitted respectively to formants $F_1$, $F_2$, and $F_3$ obtained on vowels segmented with two different processing \textit{Methods}, controlled for \textit{Gender} and \textit{Vowel} category (see section~\ref{sec:analysis}).}
    \label{tab:anova_fixed}
    \end{center}
\end{table*}

%

\begin{table*}
    \footnotesize
    \centering
    \begin{adjustbox}{width=0.75\textwidth}
        \begin{tabular}{llrrrrrl}
              \toprule
        \textbf{Model} & \textbf{Deletion} & \textbf{npar} & \textbf{loglik} & \textbf{AIC} & \textbf{LRT} & \textbf{Df} & $Pr(>\chi^2)$ \\
              \midrule
        $F_1$ & $<$none$>$   &   18 & -182288 & 364613 &     -   & - & - \\
              & $(1|S\mathord{:}(G\mathord{:}P))$  &   17 & -194498 & 389030 & 24419.6 & 1 & 0.00000 *** \\
              & $(1|G\mathord{:}P)$      &   17 & -182288 & 364611 &     0.0 & 1 & 0.90799     \\
              & $(1|P)$        &   17 & -182290 & 364614 &     2.8 & 1 & 0.09284     \\
              \midrule
        $F_2$ & $<$none$>$   &   18 & -166592 & 333220 &     -   & - & - \\
              & $(1|S\mathord{:}(G\mathord{:}P))$  &   17 & -173107 & 346248 & 13030.3 & 1 & 0.00000 *** \\
              & $(1|G\mathord{:}P)$      &   17 & -166592 & 333218 &     0.0 & 1 & 0.8749      \\
              & $(1|P)$        &   17 & -166592 & 333219 &     0.1 & 1 & 0.3931      \\
              \midrule
        $F_3$ & $<$none$>$   &   18 & -187973 & 375982 &     -   & - & - \\
              & $(1|S\mathord{:}(G\mathord{:}P))$  &   17 & -209152 & 418338 & 42358.0 & 1 & 0.0000  *** \\
              & $(1|G\mathord{:}P)$      &   17 & -187973 & 375980 &     1.0 & 1 & 0.4542      \\
              & $(1|P)$        &   17 & -187973 & 375980 &     0.0 & 1 & 1.0000      \\
              \bottomrule
        \end{tabular}
    \end{adjustbox}
    \caption{ \centering  ANOVA-like table for random effects obtained by single term deletions of the \textit{Speaker} ($S$), \textit{Gender} ($G$), and \textit{Period} ($P$) factors, for the models fitted on each formant ($F_1$, $F_2$, $F_3$). Values of the likelihood ratio test (LRT) are reported for each deleted term, compared to the full model (see section~\ref{sec:analysis}). }
    \label{tab:anova_random}
\end{table*}

\subsection{B : Cleaned Vowels Summary Table}
See the caption of the table \ref{tab:oral_vowels_table}.

\begin{table*}[ht!]
    \footnotesize
    \begin{center}
    \begin{adjustbox}{width=0.35\textwidth}        
        \begin{tabular}{lllr}
        \toprule
        \textbf{Phoneme} &  &  & \textbf{Cleaned} \\
        \midrule
        \multirow{1}{*}{\underline{Vowels}}
         & \multirow{1}{*}{Oral}
            & [\textipa{i}]  &  443,795 \\
         &  & [\textipa{y}]  &  148,757 \\
         &  & [\textipa{e}]  &  443,329 \\
         &  & [\textipa{\o}] &   89,547 \\
         &  & [\textipa{@}]  &  285,534 \\
         &  & [\textipa{E}]  &  473,529 \\
         &  & [\textipa{\oe}]&   36,875 \\
         &  & [\textipa{a}]  &  623,003 \\
         &  & [\textipa{u}]  &  149,453 \\
         &  & [\textipa{O}]  &  191,682 \\
         &  & [\textipa{o}]  &  102,608 \\
         &  & [\textipa{A}]  &   28,202 \\
         \bottomrule
        \end{tabular}
    \end{adjustbox}
    \caption{\centering Summary table of \textbf{cleaned} oral vowel returned by $MFA$ after phonetic transcription and forced alignment of the \textbf{automatic} transcriptions ($Whisper$ [large-v3]) (see section \ref{sec:corpus}).}
    \label{tab:oral_vowels_table}
    \end{center}
\end{table*}

\subsection{C: Phone Summary Table}
See the caption of the table \ref{tab:phones_table}.
\begin{table*}
    \centering
    \footnotesize
    \begin{tabular}{lllrr}
    \toprule
    \textbf{Phoneme} &  &  & \textbf{Automatic} & \textbf{Manual} \\
    \midrule
    \multirow{1}{*}{\underline{Vowels}}
     & \multirow{1}{*}{Oral}
       & [\textipa{i}]  & 706,217 & 2,047 \\
     &  & [\textipa{y}]  & 238,829 & 766 \\
     &  & [\textipa{e}]  & 675,676 & 2,118 \\
     &  & [\textipa{\o}]  & 127,524 & 435 \\
     &  & [\textipa{@}]  &  512,990 & 1,581 \\
     &  & [\textipa{E}]  & 710,354 & 2,169 \\
     &  & [\textipa{\oe}]  & 52,340 &  375 \\
     &  & [\textipa{a}]  & 923,020 &  2,754\\ 
     &  & [\textipa{u}]  & 232,540 & 672 \\
     &  & [\textipa{o}]  & 147,186 & 416 \\
     &  & [\textipa{O}]  & 262,739 & 786 \\
     &  & [\textipa{A}]  & 45,190 & 140 \\
    \cmidrule(l){2-5}
    
     & \multirow{1}{*}{Nasal}
       & [\textipa{\~a}]  & 421,404 & 1,267 \\
     &  & [\textipa{\~O}]  & 256,564 & 770 \\
     &  & [\textipa{\~E}]  & 184,927 & 551 \\
    
    \midrule
    
    \multirow{1}{*}{\underline{Consonants}}
     & \multirow{1}{*}{Plosive}
       & [\textipa{p}]  & 462,424 & 1,361 \\
     &  & [\textipa{t}]  & 627,664 & 1,847 \\
     &  & [\textipa{k}]  & 450,392 & 1,350 \\
     &  & [\textipa{b}]  & 122,218 & 349 \\
     &  & [\textipa{d}]  & 539,279 & 1,738 \\
     &  & [\textipa{g}]  & 58,156 & 172 \\
     &  & [\textipa{c}]  & 122,142 & 344 \\
     &  & [\textipa{\textbardotlessj}] & 4274 & 19 \\
    
    \cmidrule(l){2-5}
    
     & \multirow{1}{*}{Fricative}
       & [\textipa{f}]  & 175,399 & 562 \\
     &  & [\textipa{s}]  & 767,359 & 2,358 \\
     &  & [\textipa{S}]  & 60,638 & 149 \\
     &  & [\textipa{v}]  & 272,070 & 804 \\
     &  & [\textipa{z}]  & 115,033 & 332 \\
     &  & [\textipa{Z}]  & 216,210 & 672 \\
     &  & [\textipa{K}]  & 912,787 & 2,781 \\
    
    \cmidrule(l){2-5}
    
     & \multirow{1}{*}{Approximant}
       & [\textipa{4}]  & 59,972 & 171 \\ 
     &  & [\textipa{w}]  & 124,831 & 419 \\ 
     &  & [\textipa{j}]  & 175,202 & 503 \\
     &  & [\textipa{l}]  & 657,357 & 1,895 \\
     &  & [\textipa{L}]  & 56,563 & 168 \\
    \cmidrule(l){2-5}
    
     & \multirow{1}{*}{Nasal}
       & [\textipa{m}]  & 429,348 & 1,279 \\
     &  & [\textipa{n}]  & 275,343 & 825 \\
     &  & [\textipa{\textltailn}] & 39496 & 120 \\
     &  & [\textipa{N}]  & 849 & 3 \\
     &  & [\textipa{m}\textsuperscript{j}] & 6,990 & 21 \\
    \cmidrule(l){2-5}
       & \multirow{1}{*}{Affricates}
       & [\textipa{\textbottomtiebar{tS}}]  & 800 & 5 \\
     &  & [\textipa{\textbottomtiebar{dZ}}]  & 439 & 0 \\
     &  & [\textipa{\textbottomtiebar{ts}}]  & 302 & 1 \\
     
    \bottomrule
    \end{tabular}
    \caption{\centering Summary table of the number of phones returned by $MFA$ after phonetic transcription and forced alignment of the \textbf{automatic} ($Whisper$ [large-v3]) and \textbf{manual} transcriptions, without any cleaning at this stage (see section \ref{sec:corpus}). Note that $MFA$ uses a fine-grained phonetic transcription that includes some phonetic phenomena such as palatalization and non-standard French transcriptions; we kept its default choices.\\
    }
    \label{tab:phones_table}
\end{table*}

\subsection{D: ANOVA Tables} 
\label{sec:Appendix_D}

ANOVA tables for models presented in section~\ref{sec:vowels} for the analysis of diachronic changes in formants $F_1$ (Table~\ref{tab:anova_f1}) and $F_2$ (Table~\ref{tab:anova_f2}).

\begin{table*}
    \begin{center}
    \begin{adjustbox}{width=0.42\textwidth}
    \begin{tabular}{lrrl}
          \toprule
            \textbf{Factors} & $\chi^2$ & \textbf{Df} & $Pr(>\chi^2)$ \\
          \midrule
 $A$                 &         0.3767 & 1 &   0.53938     \\
 $G$              &        18.9247 & 1 & 1.360e-05 *** \\
 $P$              &         2.1955 & 6 &   0.90087     \\
 $V$               &      2618.7972 & 1 & < 2.2e-16 *** \\
 $A:G$          &         6.5716 & 1 &   0.01036 *   \\
 $A:P$          &         9.4499 & 6 &   0.14981     \\
 $A:V$           &         0.3280 & 1 &   0.56685     \\
 $G:P$       &         0.2253 & 6 &   0.99978     \\
 $G:V$        &        30.6383 & 1 & 3.109e-08 *** \\
 $P:V$        &        15.8983 & 6 &   0.01431 *   \\
 $A:P:V$    &        13.8443 & 6 &   0.03142 *   \\
          \bottomrule
    \end{tabular}
    \end{adjustbox}
    \caption{ 
    \centering Analysis of Deviance Table (Type II Wald $\chi^2$ tests) obtained for the fixed factors of the model fitted to $F_1$ obtained on /\textipa{a}/ and /\textipa{A}/ \textit{Vowels} ($V$), across  \textit{Age} ($A$), \textit{Gender} ($G$) and \textit{Period} ($P$) (see section~\ref{sec:vowels} for details).}
    \label{tab:anova_f1}
    \end{center}
\end{table*}

\begin{table*}
    \begin{center}
    \begin{adjustbox}{width=0.40\textwidth}
    \begin{tabular}{lrrl}
          \toprule
            \textbf{Factors} & $\chi^2$ & \textbf{Df} & $Pr(>\chi^2)$ \\
          \midrule
 $AT$           &      5.3243 & 1 &  0.021030 *   \\
 $G$       &      0.5006 & 1 &  0.479247     \\
 $V$        &   5377.1017 & 1 & < 2.2e-16 *** \\
 $AT:G$    &     11.5903 & 1 &  0.000663 *** \\
 $AT:V$     &    204.4467 & 1 & < 2.2e-16 *** \\
          \bottomrule
    \end{tabular}
    \end{adjustbox}
    \caption{ 
    \centering Analysis of Deviance Table (Type II Wald $\chi^2$ tests) obtained for the fixed factors of the model fitted to $F_2$ obtained on /\textipa{a}/ and /\textipa{A}/ \textit{Vowels} ($V$), along \textit{Apparent Time} ($AT$), controlled for \textit{Gender} ($G$) (see section~\ref{sec:vowels} for details).}
    \label{tab:anova_f2}
    \end{center}
\end{table*}

\end{document}